%% file: decompression.tex
\documentclass[conference, 10pt]{IEEEtran}
\usepackage{caption}
\usepackage{subcaption}
\usepackage[pdftex]{graphicx}
\usepackage{balance}
\usepackage{algorithmicx}
\usepackage{stfloats}
\usepackage{algpseudocode}
\usepackage{hyperref}
\newcommand{\eat}[1]{}
\begin{document}
\algdef{SE}[DOWHILE]{Do}{doWhile}{\algorithmicdo}[1]{\algorithmicwhile\ #1}%
\algnewcommand\algorithmicinput{\textbf{Input:}}
\algnewcommand\Input{\item[\algorithmicinput]}

\author{\IEEEauthorblockN{Evangelia Sitaridi\IEEEauthorrefmark{1},
Rene Mueller\IEEEauthorrefmark{2},
Tim Kaldewey\IEEEauthorrefmark{3}, 
Guy Lohman\IEEEauthorrefmark{2} and
Kenneth A. Ross\IEEEauthorrefmark{1}}
\IEEEauthorblockA{\IEEEauthorrefmark{1}Columbia University:
 \{eva, kar\}@cs.columbia.edu}
\IEEEauthorblockA{\IEEEauthorrefmark{2}IBM Almaden Research:
\{muellerr, lohmang\}@us.ibm.com}
\IEEEauthorblockA{\IEEEauthorrefmark{3}IBM Watson: tkaldew@us.ibm.com}
}

\title{Massively-Parallel Lossless Data Decompression}

\maketitle
           
\begin{abstract} 
Today's exponentially increasing data volumes and the high cost of storage make compression
essential for the Big Data industry. Although research has concentrated on
efficient compression, fast decompression is critical for analytics queries 
that repeatedly read compressed data.
While decompression can be parallelized somewhat by assigning each data block to a different process,
break-through speed-ups require exploiting the massive parallelism of modern multi-core processors 
and GPUs for data decompression \emph{within a block}.
We propose two new techniques to increase the degree of 
parallelism during decompression. The first technique exploits 
the massive parallelism of GPU and SIMD architectures. The second 
sacrifices some compression efficiency to eliminate data dependencies 
that limit parallelism during decompression.
We evaluate these techniques on the decompressor of the DEFLATE scheme, called
Inflate, which is based on LZ77 compression and Huffman encoding.  We achieve a 
2$\times$ speed-up in a head-to-head comparison with several multi-core CPU-based
libraries, while achieving a 17\,\% energy saving with comparable
compression ratios.
\end{abstract} 
\input intro.tex

\input related.tex

\input parallelism.tex

\input experiments.tex

\section{Conclusions and Future Work}
\label{sec:conc}

Here, we developed techniques within our compression framework, Gompresso, for 
massively parallel decompression using GPUs. We presented one solution for
parallelizing Huffman decoding by using parallel sub-blocks, and two techniques 
to resolve back-references in parallel.  The first technique iteratively 
resolves back-references and the second eliminates data dependencies during 
compression that will stall  parallelism among collaborating threads 
concurrently decompressing that set of sub-blocks. Gompresso decompresses 
two real-world datasets $2\times$ faster than the state-of-the-art 
block-parallel variant of zlib running on a modern multi-core CPU, while 
suffering no more than a 10\,\%  penalty in compression ratio. 
\textbf{Gompresso} also uses 17\,\% less energy by using GPUs.
Future work includes determining the extent to which our
techniques can be applied to alternative coding and context-based compression 
schemes, and evaluating their performance.

\section*{Acknowledgment}
Authors Evangelia Sitaridi and Kenneth Ross were partially supported by National
Science Foundation grant IIS-1218222 and by an equipment gift from NVIDIA
Corporation.

\balance 

\bibliographystyle{IEEEtran}
\bibliography{IEEEabrv,decompression}

\end{document}

%% file: intro.tex
\section{Introduction}
\label{sec:intro}

With exponentially-increasing data volumes and the high cost of enterprise data 
storage, data compression has become essential for reducing storage costs in the 
Big Data era. There exists a plethora of compression techniques, each having a
different trade-off between its compression ratio (compression efficiency) and 
its speed of execution (bandwidth). Most research so far has focused on the 
speed of compressing data as it is loaded into an information system, but the 
speed of decompressing that data can be even more important for Big Data 
workloads -- usually data is compressed only once at load time but repeatedly 
decompressed as it is read when executing analytics or machine learning jobs. 
Decompression speed is therefore crucial to minimizing response time of these 
applications, which are typically I/O-bound. 

In an era of flattening processor speeds, parallelism provides our best hope of 
speeding up any process. In this work, we leverage the massive parallelism 
provided by Graphics Processing Units (GPUs) to accelerate decompression. GPUs 
have already been successfully used to accelerate several other data processing
problems, while concomitantly providing a better Performance/Watt ratio than 
conventional CPUs, as well.  However, accelerating decompression on massively 
parallel processors like GPUs presents new challenges.  Straightforward 
parallelization methods, in which the input block is simply split into many, 
much smaller data blocks that are then processed independently by each processor, 
result in poorer compression efficiency, due to the reduced redundancy in the 
smaller blocks, as well as diminishing performance returns caused by per-block 
overheads. In order to exploit the high degree of parallelism of GPUs, with 
potentially thousands of concurrent threads, our implementation needs to take 
advantage of both intra-block parallelism and inter-block parallelism. For 
intra-block parallelism, a group of GPU threads decompresses the same data 
block concurrently. Achieving this parallelism is 
challenging due to the inherent data dependencies among the threads that 
collaborate on decompressing that block. 

In this paper, we propose and evaluate two approaches to address this 
intra-block decompression challenge. The first technique exploits the SIMD-like 
execution model of GPUs to coordinate the threads that are concurrently 
decompressing a data block. The second approach avoids data dependencies 
encountered during decompression by proactively eliminating performance-limiting 
back-references during the compression phase. The resulting speed gain comes at 
the price of a marginal loss of compression efficiency. We present 
\textbf{Gompresso/Bit}, a parallel implementation of an 
Inflate-like scheme~\cite{rfc1951} that aims at high decompression speed and is 
suitable for massively-parallel processors such as GPUs. We also implement 
\textbf{Gompresso/Byte}, based on LZ77 with byte-level encoding. It trades off
slightly lower compression ratios for an average $3\times$ higher decompression 
speed. 

In summary, the contributions of this paper are:

\begin{itemize}
\item A technique to achieve massive intra-block parallelism during 
      decompression by exploiting the SIMD-like architecture of GPUs. 
\item Improved intra-block parallelism by eliminating data 
      dependencies during compression at a slight cost of compression 
      efficiency. 
\item An evaluation of the impact of both techniques on compression ratio and 
      speed. 
\item Comparisons of Gompresso's decompression speed and energy efficiency on the 
      Tesla K40 GPU against several state-of-the-art multi-core CPU libraries, 
      in which 
      we show that \textbf{Gompresso/Bit} is $2\times$ faster while achieving a 
      17\,\% energy saving.
\end{itemize}

Section~\ref{sec:bg} gives background on the
essentials of the GPU architecture, and in Section~\ref{sec:rel} we discuss
related work. Section~\ref{sec:par} analyzes how Gompresso parallelizes
decompression to harvest the massive parallelization of GPUs.
Section~\ref{sec:conflstrat} focuses on the alternative dependency resolution
strategies we designed for LZ77. Section \ref{sec:exp} presents the experimental
results of tuning and comparing Gompresso against state-of-the-art parallel CPU
libraries. Finally, in Section~\ref{sec:conc} we summarize our conclusions and
suggest some interesting directions for future work. A shorter version of the paper
to appear as is\cite{icpp2016}.


%% file: related.tex
\section{Background and Related Work}
\label{sec:bg}
\if
In this section, we provide a brief description of the relevant background 
on data compression, based upon the LZ77~\cite{lz77} scheme combined with entropy 
coding. We also provide a short introduction to relevant aspects of modern GPU
architectures and discuss related work in parallelizing decompression.
\fi

\subsection{LZ77 Compression} 
\label{subsec:lz77bg}
\begin{figure} 
\centering
\includegraphics[width=1.0\linewidth]{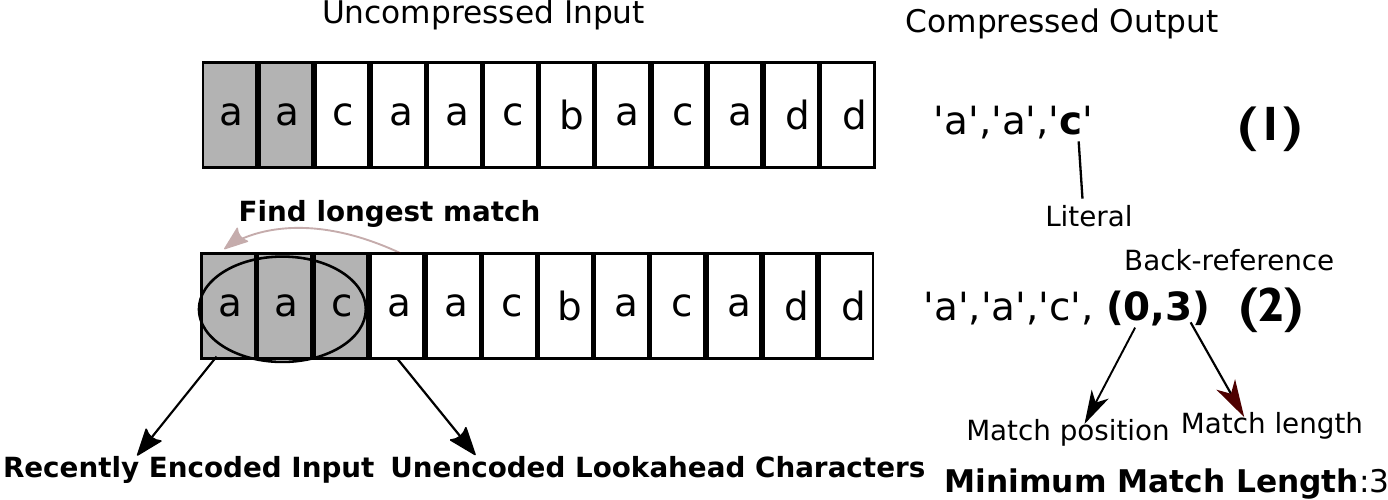}
\caption{Illustration of LZ77 compression: (1) Literal is emitted 
because there was no match for `c'. (2)  Back-reference is emitted 
for a match on `aac'.}
 \label{fig:example}
\end{figure} 
A common type of data compression replaces frequently-occurring sequences with
references to earlier occurrences. This can be achieved by maintaining a
dictionary of frequently-occurring patterns, such as in the LZ78~\cite{lz78} and
LZW~\cite{lzw} algorithms, or by maintaining a sliding window over the most
recent data, as in the LZ77 algorithm~\cite{lz77}. A further space reduction can be achieved by
encoding individual characters or dictionary symbols as variable-length code
words. This so-called \emph{entropy encoding} assigns code words with fewer bits to more
frequently occurring symbols. Popular entropy encoding schemes are Huffman
\cite{huffmancoding} or Arithmetic \cite{arithmeticcoding} coding.

Compression schemes typically combine a dictionary-based technique and entropy coding.
In this paper, we study a variant of the popular DEFLATE~\cite{rfc1951}
compression scheme, which is used in the gzip, ZIP, and PNG file formats. 
More precisely, we focus on the decompression process, called Inflate. 
DEFLATE uses the LZ77 dictionary scheme followed by Huffman coding. The dictionary in LZ77 is
a sliding window over the recently processed input. The LZ77 compressor produces
a stream of token symbols, in which each token is either a \textbf{back-reference}
to a position in the sliding window dictionary, or a \textbf{literal} containing a
sequence of characters, if that sequence does not appear in the sliding window.
A back-reference is represented as a tuple with the offset and the length of a match
in the sliding window. Figure~\ref{fig:example} illustrates both token types in
a simple example.  In the first step, the sequence `aa' has already been
processed and is in the sliding window dictionary. `caac\ldots' is the input to be
processed next. Since the window does not contain any sequence starting with
`c', LZ77 emits a literal token `c' and appends `c' to the window. The sequence
to be processed in the subsequent step is `aacb\ldots`. `aac' is found
in the window at position 0. The match is 3 characters long, hence, LZ77 emits
back-reference token (0,3).

The resulting stream of literal and back-reference tokens are then converted into
sequences of codewords by an entropy coder. DEFLATE uses Huffman coding, which
yields code words with varying bit-lengths. We also consider entropy encoding
that operates at the level of bytes rather than bits. This sacrifices 
some compression efficiency for speed. Existing dictionary-based compression
schemes that use byte-level coding are LZRW1 \cite{lzrw}, Snappy \cite{snappy}, and LZ4 \cite{lz4}.
We refer to the implementation using bit-level encoding as \textbf{Gompresso/Bit}. Similarly, 
the entropy encoder in \textbf{Gompresso/Byte} operates at the byte level.

\subsection{GPU Background} 
\label{subsec:gpubg}

We use the NVIDIA CUDA terminology and programming model; however, our work can be
extended to other standards, such as OpenCL. GPU-accelerated applications are
implemented as a number of  \emph{kernels}. A kernel is a function that is
invoked by the host CPU and is executed on the GPU device. A kernel function 
is executed by a number of \emph{thread-groups}. 
\footnote{In CUDA, thread-groups are called \emph{thread blocks} and in OpenCL \emph{work groups}. We use
the term ``group'' instead of ``block'' to avoid confusion with the term \emph{data blocks}.} 
A thread-group is further divided into smaller units of 32 threads,
called a \emph{warp}. The threads in a given warp execute all the same
instructions in lock step. Understanding this seemingly minor architectural 
detail is essential for our first data dependency resolution technique, 
described in Section~\ref{sec:par}. The execution model of a warp is essentially equivalent to a
32-way single-instruction, multiple-data (SIMD) architecture that allows branch instructions. 
In the presence of branches, the threads in a warp may have to 
follow different execution paths (diverge), based upon the outcome of the branch condition on each thread. Due to the lock-step execution of the threads within a warp, however, these different 
paths will be serialized. Thus, for better performance, 
ideally threads in a warp should not diverge and instead follow the same execution path.

Because of the significance of warps as the unit of execution, GPUs provide
several instructions that allow threads within a warp to exchange data and reach
consensus. We will use the following two instructions in this paper: The
\verb+ballot(b)+ instruction combines a binary voting bit $b_i$ from each
thread $i$ and returns them to threads as a 32-bit value $b_{31} 2^{31} + \cdots
+ b_1 2 + b_0$ that represents the individual votes. The ``shuffle'' \verb+shfl(v,i)+
instruction broadcasts the value $v$ of thread $i$ to all other threads of the
warp. 

\subsection{Related Work}
\label{sec:rel}

Although there are numerous compression schemes, we focus in this section on just the 
parallelization attempts of the best-known compression schemes.

\paragraph{Parallel CPU Implementations} 

A parallel implementation for CPUs of gzip compression in the pigz
library~\cite{pigz} achieves a linear speed-up of compression with the number of CPU cores. 
Decompression in pigz, however, has to be single-threaded because of its variable-length blocks.
Another CPU compression library, pbzip \cite{pbzip2}, parallelizes the set of algorithms
implemented by the bzip2 scheme. The input is split into data blocks that can be
compressed and decompressed in parallel. As already described in the Introduction, this inter-block parallelism alone is insufficient and results in poor performance on GPUs.

\paragraph{Hardware-Accelerated Implementations}

Parallelizing compression schemes within a block is a bigger challenge for 
massively-parallel processors. For example, the GPU implementation of bzip2 did
 not improve performance against the single-core CPU bzip2 
\cite{Patel:2012:PLD}. The major bottleneck was the string sort required for the
 Burrow-Wheeler-Transform (BWT) compression layer. Future compressor
implementations could be accelerated by replacing  string sort with suffix array
construction \cite{Deo:2013:PSA:2442516.2442536,Edwards201410,Wang2015}.

Most research has focused on accelerating
compression, rather than decompression~\cite{ozsoy11}. Here, we
address the thread dependencies that limit the parallelism of the LZ77
decompression. In our implementation each thread 
writes multiple back-reference  characters at a time, avoiding the high
per character cost. A parallel algorithm for LZ decompression, depending on the
type of data dependencies, does not guarantee efficient GPU memory access\cite{878174}.
Huffman encoding is typically added to improve the compression 
ratio\cite{OzsoySC14}.  However, decoding is hard to parallelize
because it has to identify codeword boundaries for variable-length coding schemes. 
Our parallel decoding method splits data blocks into smaller 
sub-blocks to increase the available parallelism. We trade-off a little of compression efficiency
 but only make only one pass over the encoded data. Alternative parallel
decoding algorithms do not affect the compression ratio but they require multiple passes 
to decode the data for BWT decompression: A first pass to determine the 
codeword boundaries  and a second for the actual decoding 
\cite{Edwards201410}.

Simpler compression schemes have been
 implemented on GPUs in the context of a database system \cite{Fang}, but while
 these algorithms achieve good compression ratios for database columns, they 
are not efficient for Big Data workloads that might be unstructured. FPGAs and 
custom hardware have also been used to accelerate compression, resulting in high 
speed-ups \cite{xilinx,Abdelfattah}. However, these hardware devices have very 
different characteristics and constraints than GPUs, so their parallelization 
techniques generally aren't  applicable to GPUs.


%% file: parallelism.tex
\begin{figure*}[ht!] 
\centering
\includegraphics[width=0.75\linewidth]{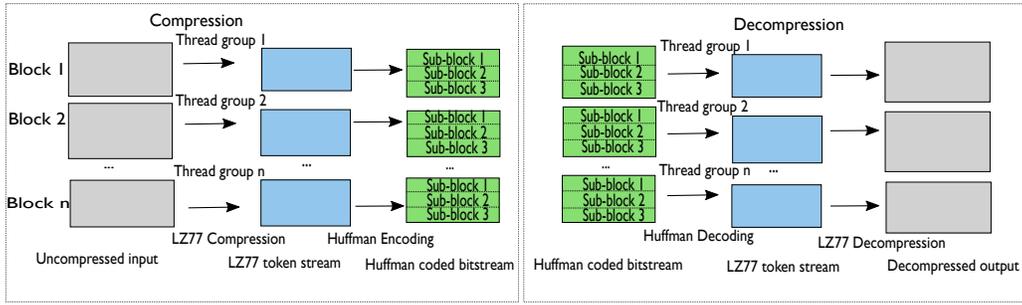}
\caption{\textbf{Gompresso} splits the input into equally-sized blocks, which are then
LZ77-compressed independently and in parallel. In \textbf{Gompresso/Bit}, the 
resulting token streams are further split into equal-sized sub-blocks 
and Huffman-encoded. The inverse process follows for decompression.}
\label{fig:gomp_overview}
\end{figure*}  

\section{\textbf{Gompresso} Overview}
\label{sec:par}

In this section, we provide an overview of \textbf{Gompresso}, which exploits parallelism
between and also within data blocks. The most important design goal for
\textbf{Gompresso} is a high decompression speed, while maintaining a ``reasonable''
compression ratio. \textbf{Gompresso} implements both compression and decompression, and
defines its own file format.  

Figure~\ref{fig:gomp_overview} gives an overview
of the \textbf{Gompresso} compression and decompression algorithms. We first briefly outline the
parallel compression phase before describing parallel decompression, which is
the focus of this paper.

\subsection{Parallel Compression}
\label{subsec:parcomp}

In the first step, \textbf{Gompresso} splits the input into equally-sized data blocks,
which are then compressed independently and in parallel. The block size is a
configurable run-time parameter that is chosen depending on the total data size and
the number of available processing elements on the GPU. Each block is
LZ77-compressed by a group of threads using an exhaustive parallel matching
technique we described earlier~\cite{gompressogtc}. For \textbf{Gompresso/Byte}, the
pipeline ends here, and the resulting token streams are written into the output
file using a direct byte-level encoding. \textbf{Gompresso/Bit} requires an additional
step in which the tokens are encoded using a Huffman coder. Similar to DEFLATE,
\textbf{Gompresso/Bit} uses two separate Huffman trees to facilitate the encoding, 
one for the match offset values and
the second for the length of the matches and the literals themselves. Both trees
are created from the token frequencies for each block.  To facilitate parallel
decoding later on, the tokens of the data blocks are further split into smaller
sub-blocks during encoding. A run-time parameter allows the user to set the 
number of sub-blocks per data block; more sub-blocks per block increases parallelism and
hence performance, but diminishes sub-block size and hence compression ratio. 
Each encoded sub-block is
written to the output file, along with its compressed size in bits.  The
parallel decoder can determine the location of the encoded sub-blocks in the
compressed bitstream with this size information.  Finally, the Huffman trees are
written in a canonical representation~\cite{huffmancoding}.

\begin{figure}[b!] 
\centering
\includegraphics[width=1.0\linewidth]{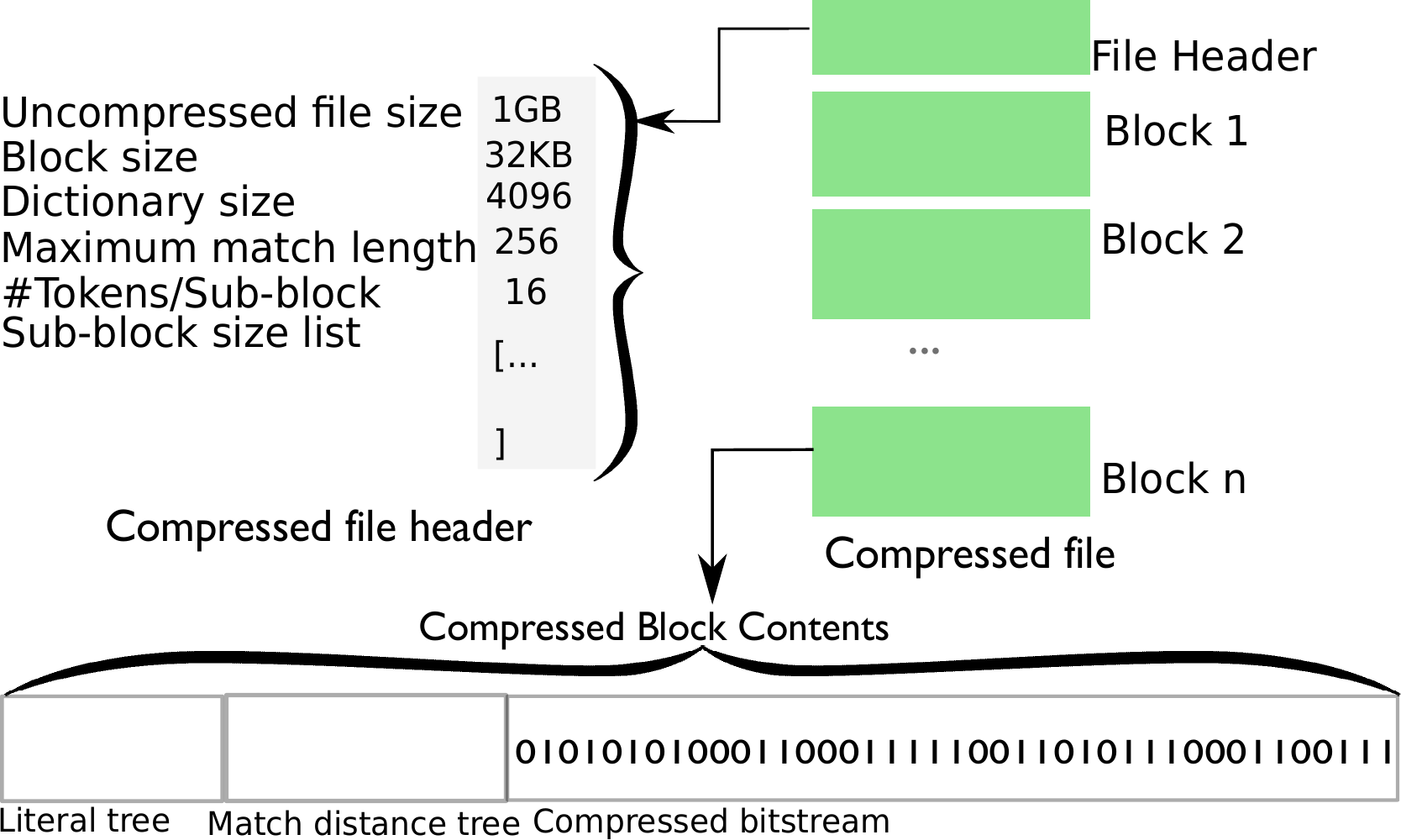}
\caption{The \textbf{Gompresso} file format, consisting of: (1) a file header, (2) a sequence
of compressed data blocks, each with its two Huffman trees (\textbf{Gompresso/Byte} does not use Huffman trees.) and encoded bitstream.}   
\label{fig:file_format}
\end{figure}

Figure~\ref{fig:file_format} shows the structure of the compressed file format
in detail.

\subsection{Parallel Decompression}
\label{subsec:pardecomp}

\textbf{Gompresso/Byte} can combine decoding and decompression in a single pass because of
its fixed-length byte-level coding scheme. The token streams can be read directly from
the compressed output. \textbf{Gompresso/Bit} uses a variable-length coding scheme for a
higher compression ratio, and therefore needs to first decode the bitstream into a stream
of tokens before proceeding with the LZ77 decompression. \textbf{Gompresso} assigns a
group of GPU threads to collaborate on the Huffman decoding and LZ77
decompression on the independently compressed data blocks. This permits an
additional degree of parallelism within data blocks.

\subsubsection{Huffman Decoding} Each thread of a group decodes a different
sub-block of the compressed data block.  The starting offset of each sub-block in
the bitstream is computed from the sub-block sizes in the file header. All
sub-blocks of a given data block decode their bitstreams using look-up tables created 
from the same two Huffman trees for that block and stored in the software-controlled, 
on-chip memories of the GPU. 
We can retrieve the original token symbol with a single lookup in each table, which is much
faster than searching through the (more compact) Huffman trees, which would introduce branches and hence divergence of the threads' execution paths.  The output of the decoder is the stream of
literal and back-reference tokens, and is written back to the device memory.  

\subsubsection{LZ77 Decompression} Each data block is assigned to a single GPU warp (32 threads
operating in lock-step) for decompression. We chose to limit the group size to one warp in order to be able to
take advantage of the efficient voting and shuffling instructions within a warp. Larger
thread groups would require explicit synchronization and data exchange via
on-chip memory. We found that the potential performance gain by the increased
degree of parallelism is canceled out by this additional coordination overhead.

\begin{figure}[hb!] 
\centering
\includegraphics[width=0.6\linewidth]{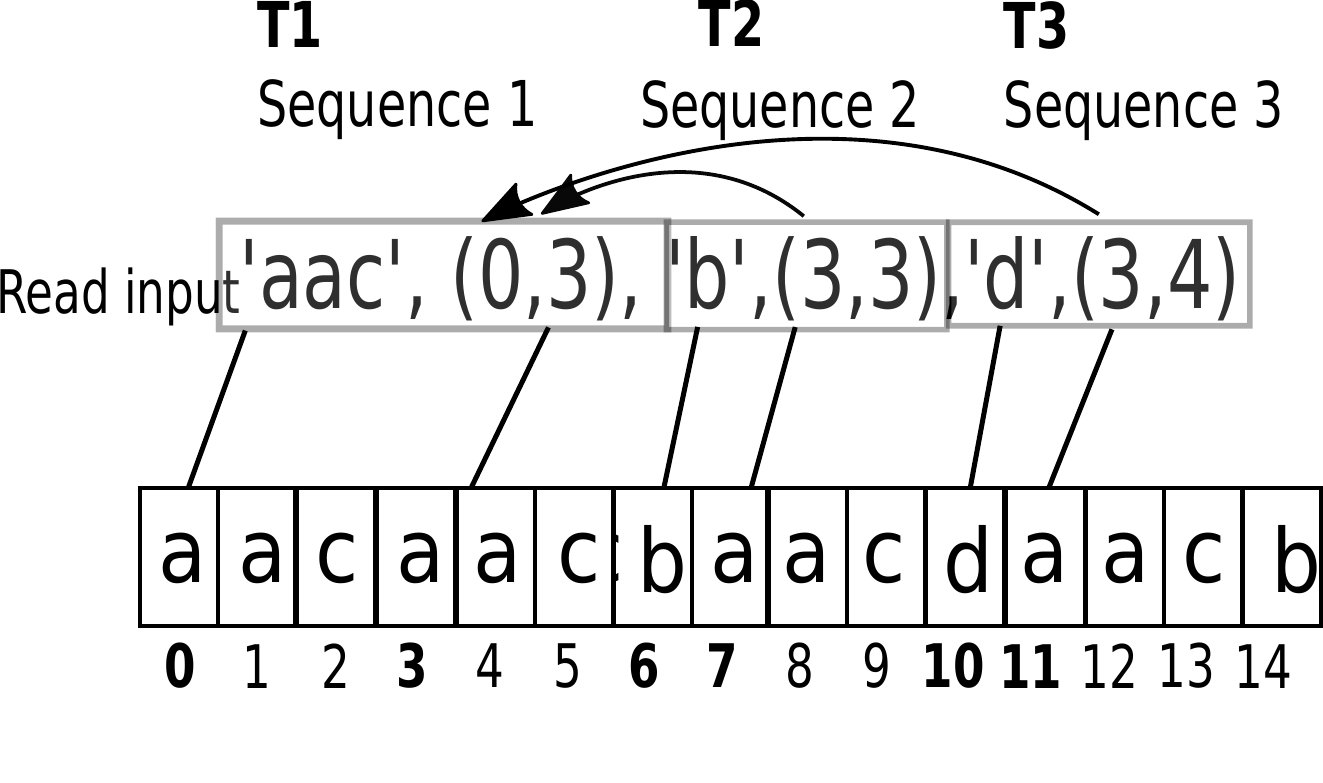}
\caption{Nested back-references: back-references in Sequence 2 and 3
depend on Sequence 1, and cannot be resolved before the output of Sequence 1
is available.}
\label{fig:ex_confl}
\end{figure}

We first group consecutive literals into a single literal string. We further require that a
literal string is followed by a back-reference and vice versa, similar to the
LZ4~\cite{lz4} compression scheme. A literal string may have zero length if
there is no literal token between two consecutive back-references. A pair
consisting of a literal string and a back-reference is called a
\textit{sequence}. We assign each sequence to a different thread (see
Figure~\ref{fig:ex_confl}). In our experiments, we found that this grouping
results in better decompression speed, since it not only assigns each thread a
larger unit of work but its uniformity suits the lock-step execution model of the
GPU. All threads in the warp concurrently alternate between executing instructions for string literals
and for back references. For each sequence, its thread performs: (a) read
its sequence from device memory and compute the start position of its string
literal, (b) determine the output position of its literal, and copy its string
literal to the output buffer, and (c) resolve and write its back-reference. 
We now describe each step in more detail:

\paragraph{Reading sequences} Each warp uses the block offset to determine
the location of the first decoded token in the device memory. Each thread in the
warp will read a different sequence (see Figure~\ref{fig:ex_confl}). The
threads then need to determine the start location of their literal strings in
the token stream. This is accomplished by computing an intra-warp exclusive
prefix sum from the literal lengths of their sequences, in order to locate the
start positions from which they can copy their literal strings.  We use NVIDIA's shuffle
instructions to efficiently compute this prefix sum without memory accesses, a common GPU technique.

\paragraph{Copying literal strings}
Next, the threads compute write positions in the decompressed output buffer.
Since all blocks, except potentially the last, have the same uncompressed size,
the threads can also easily determine the start position of their block in the
uncompressed output stream. The start position of each thread's literal string
is determined by a second exclusive prefix sum, which is then added to the start
position of the block. This prefix sum is computed from the total number of
bytes that each thread will write for its sequence, i.e., the length of its literal
string plus the match length of the back-reference. Once the source and destination
positions are determined from the two prefix sums, the threads can 
copy the literal strings from the token stream into the output buffer.

\paragraph{Copying back-references}
This is the most challenging step for parallel decompression,
because of the data dependencies between threads in a warp. These dependencies
arise when a back-reference points to another back-reference, and thus
cannot be resolved before the former has been resolved. We address these nested 
back-references in Section~\ref{sec:conflstrat}. After all the back-references have
been resolved, the warp continues with the next 32 sequences.

\section{Data Dependencies in Nested back-references}
\label{sec:conflstrat}

Before processing a back-reference, the data pointed to by this reference needs
to be available in the output. This introduces a data dependency and stalls
threads with dependent references until the referenced data becomes available.
The problem is illustrated in Figure~\ref{fig:ex_confl}. Threads T2 and T3 will
have to wait for T1 to finish processing its sequence, because they both have
back-references that point into the range that is written by T1. Resolving
back-references sequentially would produce the correct output, but would also
under-utilize the available thread resources. To maximize thread utilization, we
propose two strategies to handle these data dependencies. The first strategy
uses warp shuffling and voting instructions to process dependencies as soon as
possible, i.e., as soon as all of the referenced data becomes available. The
second strategy avoids data dependencies altogether by prohibiting construction of nested
back-references during compression.  This second approach unfortunately reduces
compression efficiency somewhat, which we will quantify experimentally in Section \ref{sec:exp}.

\begin{figure} 
\begin{algorithmic}[1]
\Function {MRR}{HWM, read\_pos, write\_pos, length}
  \State{pending $\gets$ \textbf{true}} 
          \Comment{thread has not written any output} \label{alg:pendinginit}
  \Do
    \If{pending \textbf{and} read\_pos$+$length$\leq$HWM}  \label{alg:chhwm}
      \State copy length bytes from read\_pos to write\_pos \label{alg:copy}
      \State pending $\gets$ \textbf{false} \label{alg:pendingclear}
    \EndIf
    \State votes $\gets$ ballot(pending) \label{alg:voting}
    \State last\_writer $\gets$ count\_leading\_zero\_bits(votes)
  \State HWM $\gets$ shfl(write\_pos+length, last\_writer) \label{alg:shfl2}
  \doWhile{votes$>0$} \Comment{Repeat until all threads done}
  \State \Return{HWM}
\EndFunction
\end{algorithmic}
\caption{Multi-Round Resolution (MRR) Algorithm}
\label{alg:mrr}
\end{figure} 

\subsection{Multi-Round Resolution (MRR) of Nested back-references}
\label{subsec:mrr}

Figure~\ref{alg:mrr} shows the Multi-Round Resolution (MRR) algorithm for 
iterative resolution of nested back-references, which is executed by every thread in the warp. 
We follow the GPU programming convention in which each of the variables is \emph{thread-private} unless
it is explicitly marked as locally or globally shared. The Boolean variable 
\textit{pending} is initially set on~\ref{alg:pendinginit} and is cleared once
the thread has copied its back-reference to the output (line~\ref{alg:pendingclear}). 

Before calling MRR, all threads have written their literal string from their
sequence to the output, but no thread in the warp has written a back-reference
yet. In order to determine when the referenced data becomes available, the
threads keep track of the high-water mark (HWM) position of the output that has
been written so far without gaps. A back-reference whose referenced interval is
below the HWM can therefore be resolved. In each iteration, threads that have
not yet written their output use the high-water mark (HWM) to determine whether
their back reference can be resolved (line~\ref{alg:chhwm}). If so, they copy 
the data from the referenced sequence to the output, and indicate that they 
completed their work (lines~\ref{alg:copy} and \ref{alg:pendingclear}). 

The HWM is updated at the end of each iteration. The algorithm determines the
last sequence that was completed by the warp, and sets the HWM past the highest
write position of that sequence's back-reference. The threads can determine the
last sequence without accessing shared memory by exploiting the warp-voting
instruction $\texttt{ballot}$ on the \texttt{pending} flag
(line~\ref{alg:voting}). This produces a 32-bit bitmap that contains the
\texttt{pending} states of all threads in this warp. Each thread receives this
bitmap and then counts the number of leading zeros in the bitmap in order to
determine the ID of the \texttt{last\_writer} thread that completed
the last sequence. A subsequent warp-shuffle instruction broadcasts the new HWM
computed by the \texttt{last\_writer} thread to all other threads in the warp
(line~\ref{alg:shfl2}). The iteration completes when all threads have processed
their back-references. 

\begin{figure}[hb!] 
  \centering
  \includegraphics[width=0.75\linewidth]{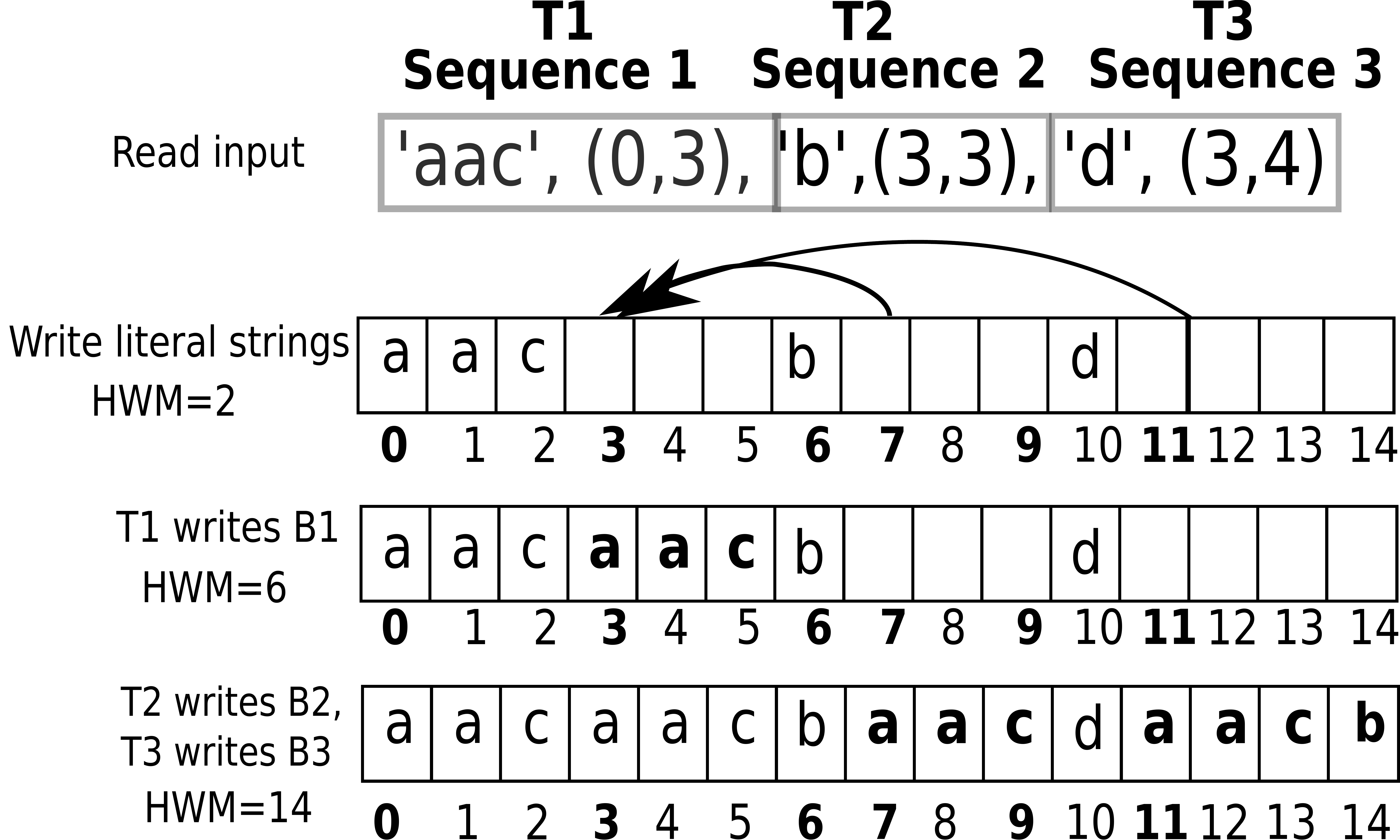}
  \caption{Multi-Round Resolution (MRR) execution}
  \label{fig:ex_mrr}
\end{figure} 

Figure~\ref{fig:ex_mrr} illustrates the execution of MRR, for the set of 3
sequences from Figure~\ref{fig:ex_confl}. 
Initially, all threads write in parallel their string of literals.
In the next step, T1 copies the back-reference of Sequence 1. In the last step,
after Sequence 1 has been processed, the dependencies of T2 and T3 are satisfied,
so both threads can proceed to copy their back-references.

At least one back-reference is resolved during each iteration which guarantees
termination of the algorithm. The HWM increases strictly monotonically. The
degree of achievable parallelism depends on nesting of back-references. As soon
as the referenced ranges falls below the HWM they can be resolved simultaneously.
Back-references that do not depend on data produced by other back-references from the
same warp can be resolved in one round leading to maximum parallelism of the warp. 
In the worst-case scenario all but one back-reference depends on another back-reference 
in the same warp. MRR then leads to sequential execution. The next
section describes a strategy that avoids this scenario.

\begin{figure} 
\begin{algorithmic}[1]
\State pos $\gets 0$
\While{pos$<$blocksize}
\State warpHWM $\gets$ pos \label{alg:deupdate}
\State s $\gets\,0$
\State literal\_str $\gets$ ``''
\While{s $<32$}
\State match $\gets$ find\_match\_below\_hwm(dict, input, \\ 
       $\qquad\qquad\qquad\qquad\qquad\qquad\qquad\qquad$ warpHWM)
\label{alg:dematchcall}
\If{	match found }
\State emit\_sequence$\big(($literal\_str, match$)\big)$ \label{alg:deemitseq}
\State update\_dictionary\_with\_backref(dict, match)
\State pos $\gets$ pos $+$ match.length
\State s $\gets$ s $+$ 1
\State literal\_str $\gets$ ``''
\Else
\State b $\gets$ get next byte from input
\State literal\_str $\gets$ literal\_str $|$ b \label{alg:deaddliteral}
\State update\_dictionary\_with\_literal\_byte(dict, b) \label{alg:deaddliteraldict}
\State pos $\gets$ pos $+\,1$
\EndIf
\EndWhile
\EndWhile
\end{algorithmic}
\caption{Modified LZ77 compression algorithm with Dependency Elimination (DE)}
\label{alg:dedeflate}
\end{figure} 

\begin{figure}[hbt!] 
\centering
\includegraphics[width=0.7\linewidth]{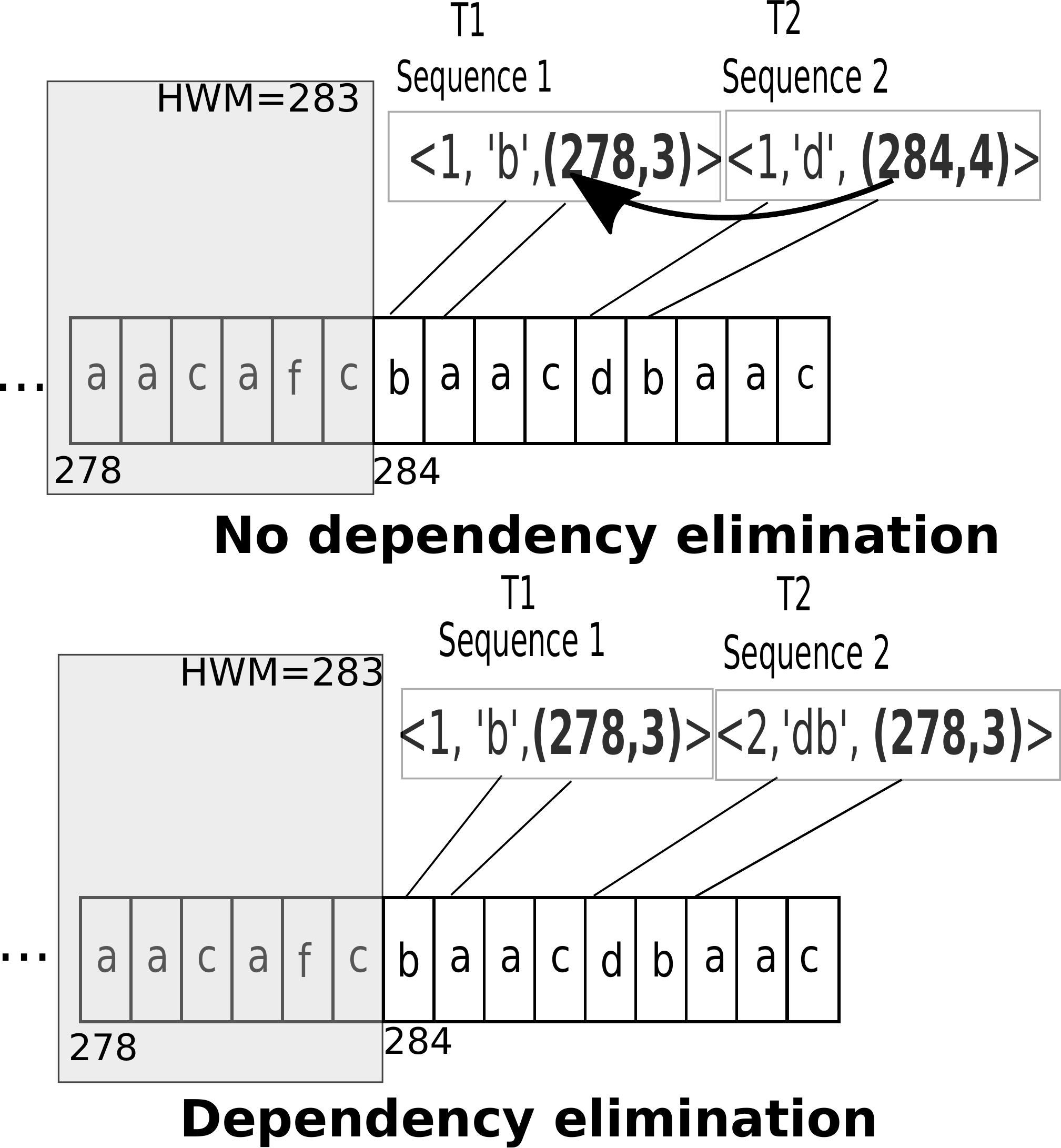}
\caption{Resulting token stream without and with dependency elimination (DE)}
 \label{fig:ex_conflde}
\end{figure} 

\subsection{Dependency Elimination (DE)}
\label{subsec:de}

In this strategy, we trade off some compression efficiency to avoid MRR's run-time cost of iteratively
detecting and resolving dependencies during decompression. During compression, we prohibit
nested back-references that would create data dependencies within the same warp. 
This does not eliminate \emph{all} nested back-references, only those
that would depend on other back-references \textit{within the same warp}. 
Prohibiting these same-warp back-references generally results in a slightly lower
compression ratio and more effort during compression, due to the additional checking and bookkeeping. 
As we will show in Section~\ref{sec:exp}, the degradation in
compression ratio and compression speed is acceptable. In return, however, we
get a 2--3$\times$ gain in decompression speed.

Dependency elimination works as follows: For every group of 32 sequences that
will eventually be decompressed by the same warp of threads, we only look for
dictionary matches below a certain warp high-water mark (warpHWM). By choosing
the warpHWM to be the cursor position in the input that has been completed
previously by the warp, we avoid
back-references that would otherwise lead to data dependencies.
Figure~\ref{alg:dedeflate} shows the modified LZ77 compression algorithm. The
warpHWM is updated only after a group of 32 sequences have been completely processed
(line~\ref{alg:deupdate}).  Threads that cooperate in the compression perform
the string matching in parallel in \texttt{find\_match\_below\_hwm}
(line~\ref{alg:dematchcall}).  They only look for a match below the current warpHWM. 
If no match is found, the next input byte is added to the literal string
(line~\ref{alg:deaddliteral}) and to the dictionary
(line~\ref{alg:deaddliteraldict}). Otherwise, if a match is found, the thread closes and emits 
the output sequence comprising the current literal
string and the found match as a back-reference (line~\ref{alg:deemitseq}). Then
the dictionary is updated with the found match.  The variable ``pos'' keeps
track of the cursor position in the processed input. 
Figure~\ref{fig:ex_conflde} illustrates the algorithm with an example. The 
dependency of T2 on T1 is avoided by choosing a shorter match in the
back-references for Sequence 2. 

Since our \textbf{Gompresso} work is focused on decompression, our implementation of the
compressor is not as highly optimized as the most commonly used data compression
libraries. We decided to implement the DE algorithm in the LZ4 compression
library (CPU-only) \cite{lz4} in order to measure the impact that the dependency
elimination has on compression speed and the resulting compression ratio. In
addition to the DE algorithm itself, we also had to implement the logic for
\texttt{find\_match\_below\_hwm()} (line~\ref{alg:dematchcall}) by
modifying the match-finding component in the LZ4 library so that it only
returns matches below a certain HWM.  To find matches, the compressor of the LZ4 library uses a
hash table, a common choice for single-threaded implementations of
LZ-based compression. The key in the hash table is a string of
three bytes (trigram). The value is the most recent position in the input in which
that trigram was encountered. This most recent position needs to be compared with
the warpHWM. We modified the existing hash replacement policy to replace an
occurrence with a more recent one only if the original entry is at more than
some number of bytes behind the current byte position. We use a constant value
for this ``minimal staleness'', which we determined experimentally. By testing
different values ranging from 64--8\,K on different datasets, we determined that 1\,K
results in the lowest compression ratio degradation.


%% file: experiments.tex
\section{Experimental Evaluation}
\label{sec:exp}

\begin{figure*}[ht!] 
\centering
\begin{subfigure}{.32\textwidth} 
 \includegraphics[width=1.0\linewidth]{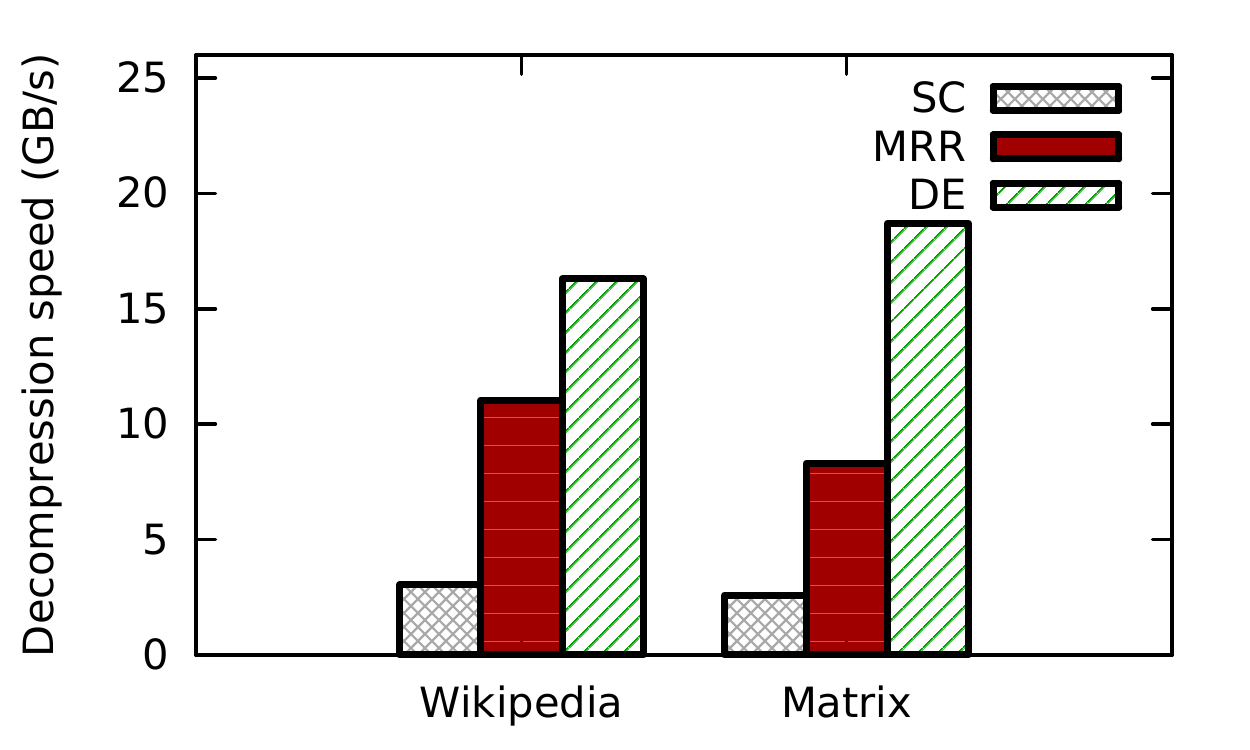}
 \subcaption{}
 \label{sfig:conflstrat}
\end{subfigure} 
\begin{subfigure}{.34\textwidth} 
\includegraphics[width=1.0\linewidth]{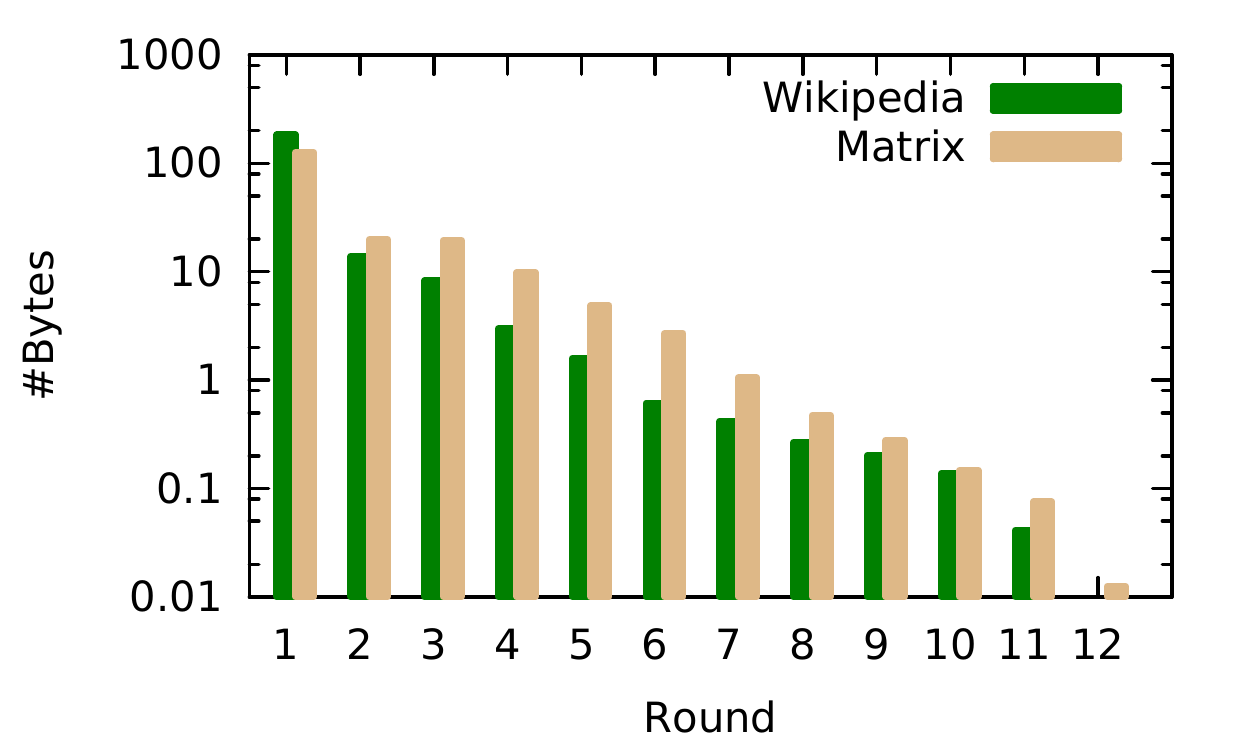}
\subcaption{}
   \label{sfig:mrrbytesperround}
  \end{subfigure} 
 \begin{subfigure}{.32\textwidth} 
  \centering
\includegraphics[width=1.0\linewidth]{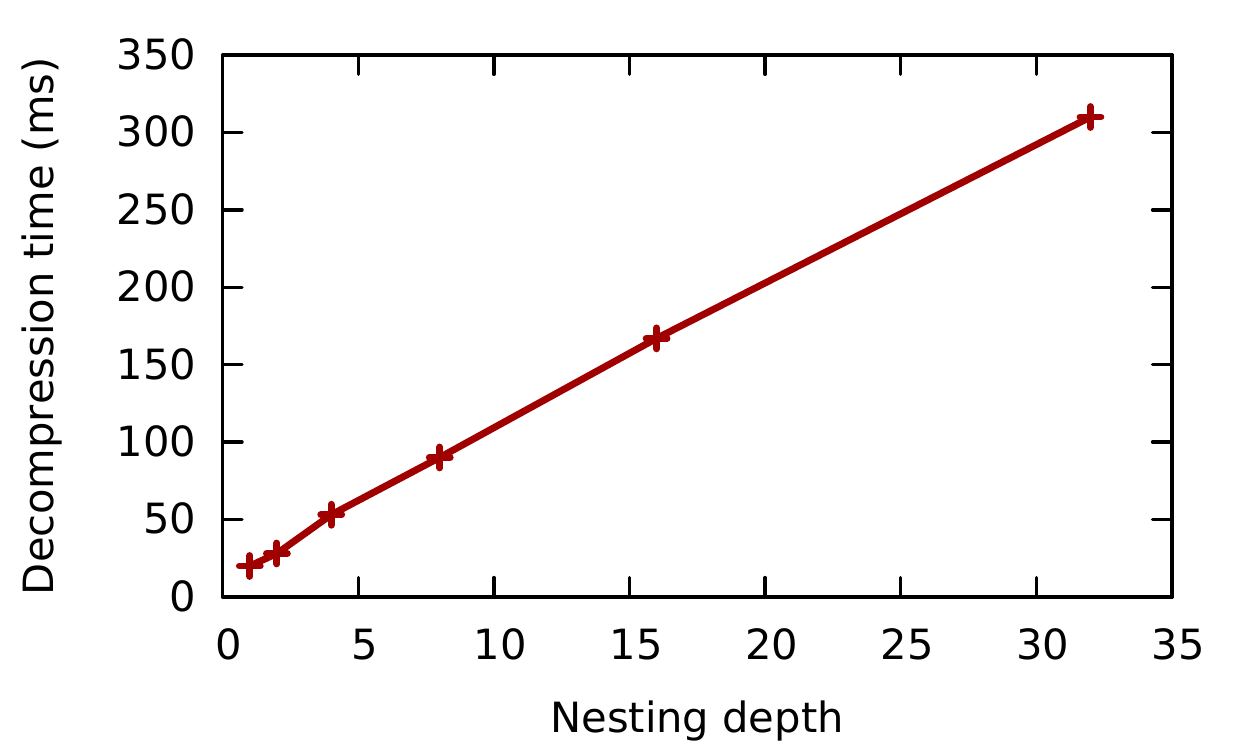}
   \subcaption{}
   \label{fig:mrrnested}
  \end{subfigure} 
\caption{(a) Decompression speed of \textbf{Gompresso/Byte} (data transfer cost not included),
using different dependency resolution strategies for the two datasets. (b) 
Number of bytes processed on each round of MRR. (c) Decompression speed 
of MRR as a function of the number of resolution rounds, for an artificially generated
dataset.}
\label{fig:cstrat}
\end{figure*} 

We evaluate \textbf{Gompresso} using two different datasets. The first is a 1\,GB XML
dump of the English Wikipedia \cite{enwik}. The second dataset is the "Hollywood-2009" sparse
matrix from the \emph{University of Florida Sparse Matrix Collection}, stored as a
0.77\,GB CSV file Matrix Market file \cite{hollywoodmm}. Both sets are highly compressible. For comparison, the gzip tool achieves a
compression ratio of 3.09:1 for the former and 4.99:1 for the latter, using the
default compression level setting (--6). The performance measurements are
conducted on a dual-socket system with two Intel E5-2620 v2 CPUs, $2\times 6$
cores running 24 hardware threads. We add an NVIDIA Tesla K40 with 2,880
CUDA cores to the system for the GPU measurements.  The device is connected via a
PCI Express (PCIe) 3.0 x16 link with a nominal bandwidth of 16\,GB/sec in each direction. We report bandwidth numbers that include PCIe transfers. In cases in which the
PCIe bandwidth becomes the bottleneck, we report the bandwidth with input and
output data residing in the GPU's device memory. ECC is turned on in our measurements. We determine the decompression bandwidth as the ratio of the size of the \emph{uncompressed data} over the
total processing time. Unless otherwise noted, we are using a data block size of
256\,KB and a sliding window of 8\,KB. For compression, we look at the next 64
bytes in the input for each match search in the 8\,KB window. To facilitate
parallel Huffman decoding in \textbf{Gompresso/Bit}, we split the sequence stream into
sub-blocks that are 16 sequences long.

\subsection{Performance Impact of Nested back-references} 

We first focus on just the LZ decompression throughput of \textbf{Gompresso/Byte}, i.e., with no
entropy coding, for different resolution strategies in
Figure~\ref{sfig:conflstrat}. \emph{Sequential Copying (SC)} is our baseline, in
which threads copy their back-references in a sequential order without
intra-block parallelism. The figure shows that Dependency Elimination (DE) is the fastest
strategy for decompression. It is at least 5$\times$ faster than SC. We place the
compressed input and the decompressed output in device memory in this setup, and
ignore PCIe transfers. The figure shows that the decompression throughput is
higher than the theoretical maximal bandwidth of the PCIe link. As expected, Multi-Round Resolution 
(MRR) performs better than SC due to the higher degree of parallelism, while DE 
out-performs MRR because it achieves an even higher degree of parallelism. 

Figure~\ref{sfig:mrrbytesperround} shows the average number of bytes that are
resolved from back-references in each round. For example, for round 2, we sum the number of bytes copied by the active threads in the second round divided by the number of MRR iterations
executed for a dataset. The lower performance
of MRR was surprising, given that we observed relatively few bytes processed
after the first round. However, what limits performance is the number
of rounds. For the Wikipedia dataset, the average number of resolution rounds is
around 3, and for the Matrix dataset, 4.

To better understand the performance impact of multiple passes, we created a
collection of artificial 1\,GB datasets that induce a specified depth of
back-reference nesting. We generate each dataset such that it leads to the
desired depth. The general idea is as follows: we repeat a 16-byte string with a
one-byte change occurring in an alternating fashion at the first and last byte
position. We chose the length of 16 to be close to the average
back-reference match length in the two real datasets used in our evaluation. A separator byte, chosen from a disjoint set of bytes, is used to
prevent accidental and undesired matches that cross different instances of the
repeating string. 
\begin{figure}[ht!] 
\centering
\includegraphics[width=0.7\linewidth]{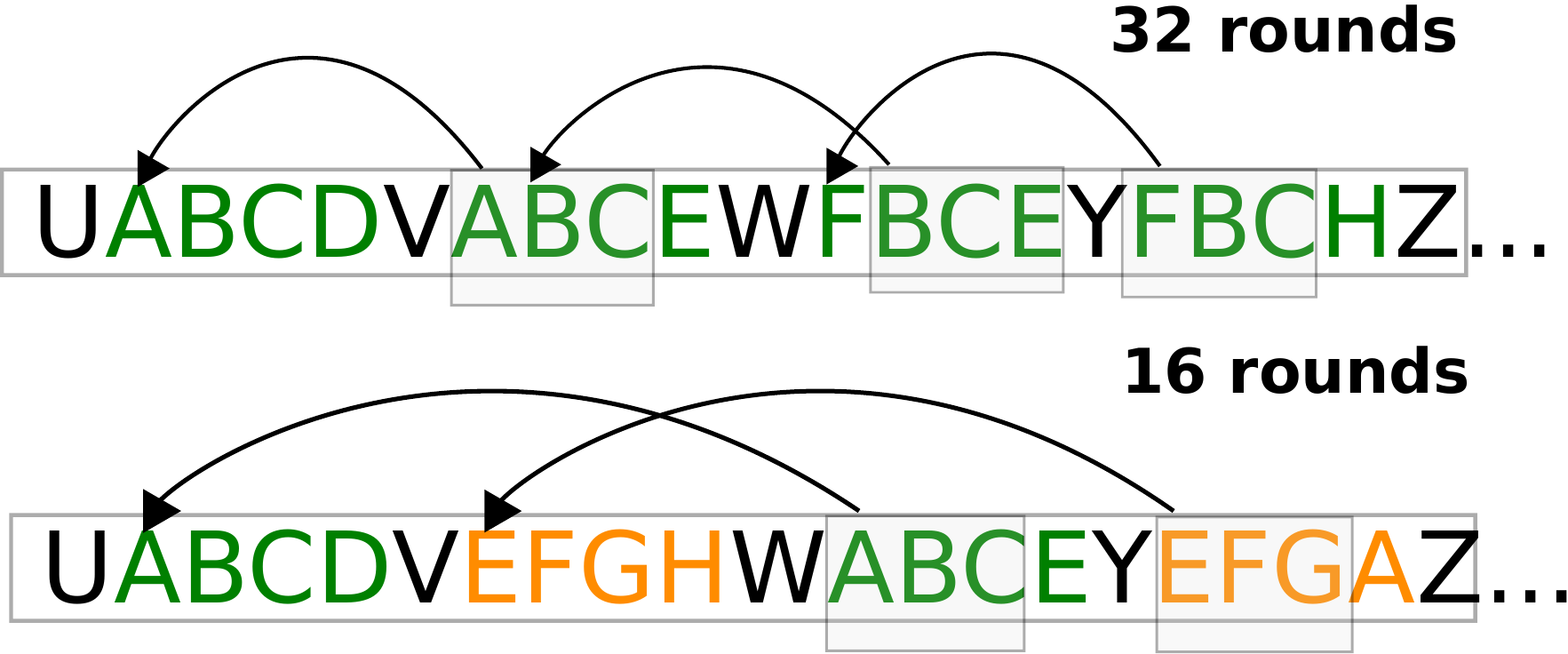}
\caption{ Series of sequences inducing 32 and 16 rounds of resolution.}
\label{fig:32nesting}
\end{figure} 

Figure~\ref{fig:32nesting} illustrates how sequences of nested
back-references are created. We show two small examples for strings of length
four, rather than 16 bytes, for space reasons. The separator bytes are printed in
black, while the repeating string is shown in green and orange colors. The arrows
show the dependencies in the MRR algorithm. LZ decompression of the dataset
shown on top in Figure~\ref{fig:32nesting} will incur data dependencies of all
32 threads in the warp except the first, whose dependency does not cause a
stall because it points to the data that was processed by this warp previously.
The nesting depth in a warp is 32, so completing the resolution requires 32 rounds.
In order to generate datasets with a smaller nesting depth, we alternate
multiple distinct repeated strings. For example, two repeated strings result in
depth 16, four repeated strings in depth 8, and so on. For a depth of 16, in each
round, two back-references are copied, one for each repeated string. These two
strings are marked in green and orange in the lower example in
Figure~\ref{fig:32nesting}.  Figure~\ref{fig:mrrnested} shows the decompression
time for different nesting depths. The decompression time increases sharply until about 16 rounds. The primary reason for the slower performance of MRR is that all
threads in a warp have to wait until the entire warp's back-references have been
resolved. Threads that resolve on the first round will be underutilized while
other threads do work in subsequent passes. 

We also implemented an alternative variant of MRR that wrote nested back-references to device memory during each round. Each round is performed in a separate kernel. Later passes read unresolved back-references and all threads in a warp can be doing useful work. Because of the overhead of writing to and reading from memory, together with the
increased complexity of tracking when a dependency can be resolved, the
alternative variant did not improve the performance of MRR.

\begin{figure} 
\centering
\includegraphics[width=0.8\linewidth]{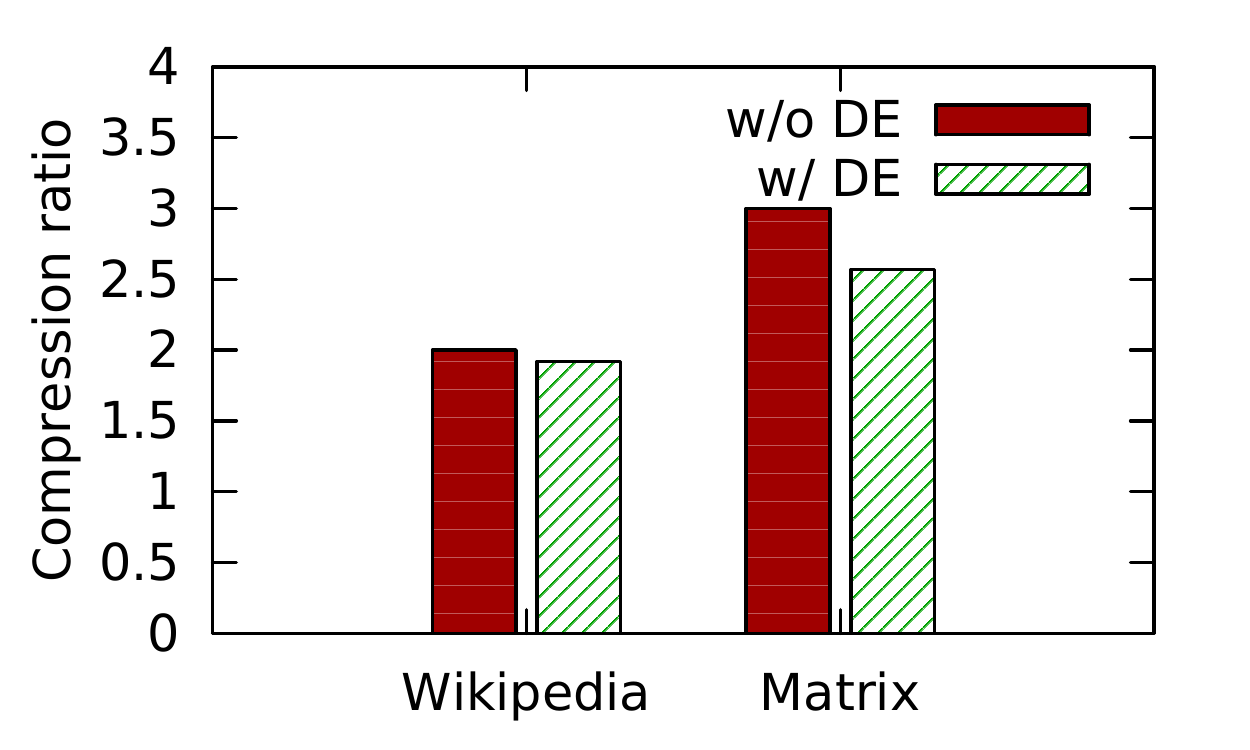}
\includegraphics[width=0.8\linewidth]{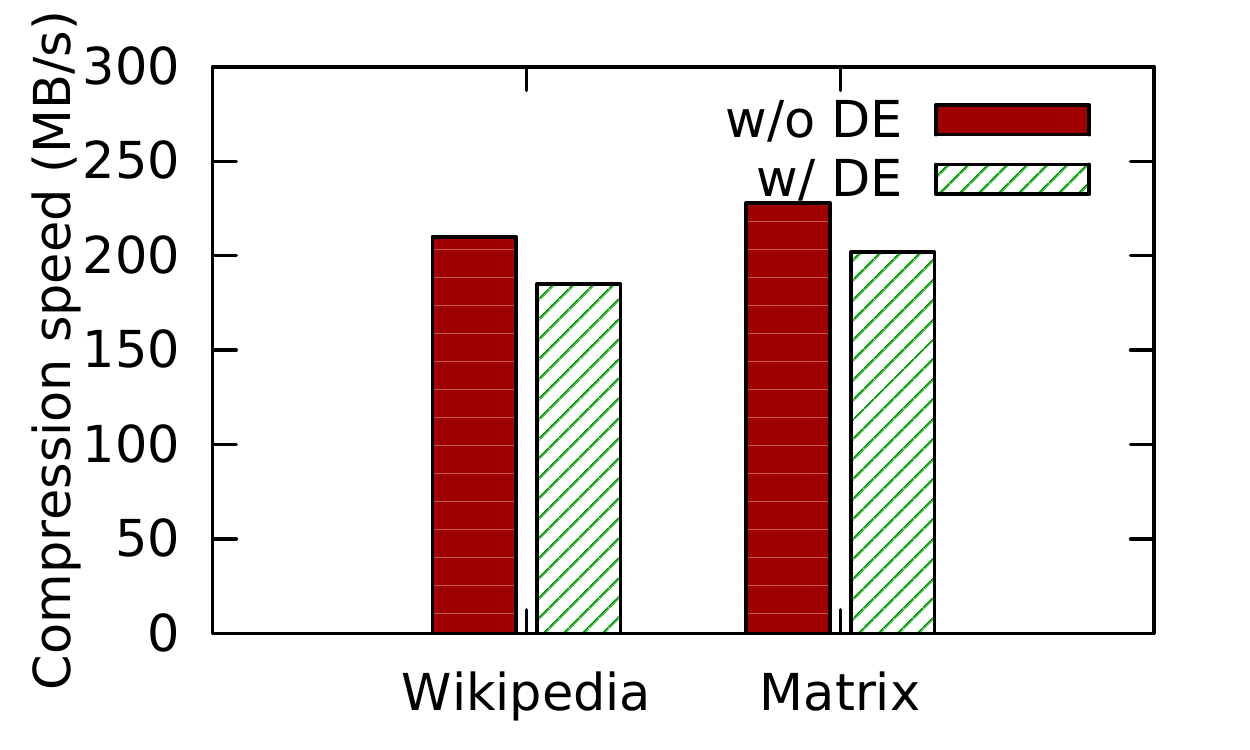}
\caption{Degradation in compression ratio and compression speed}
 \label{fig:overhead}
\end{figure}  

\subsection{Impact of DE on Compression Ratio and Speed}
Figure~\ref{fig:overhead}  shows the degradation in compression ratio and compression speed when eliminating dependencies using the Dependency Elimination (DE) algorithm we implemented by modifying the LZ4 library. The maximum degradation is 13\,\% in compression speed and 19\,\% in compression ratio, which is acceptable when we are aiming at fast decompression. In the remaining experiments, we use the DE method for decompression.

\begin{figure}[hbt!] 
\centering
  \includegraphics[width=0.8\linewidth]{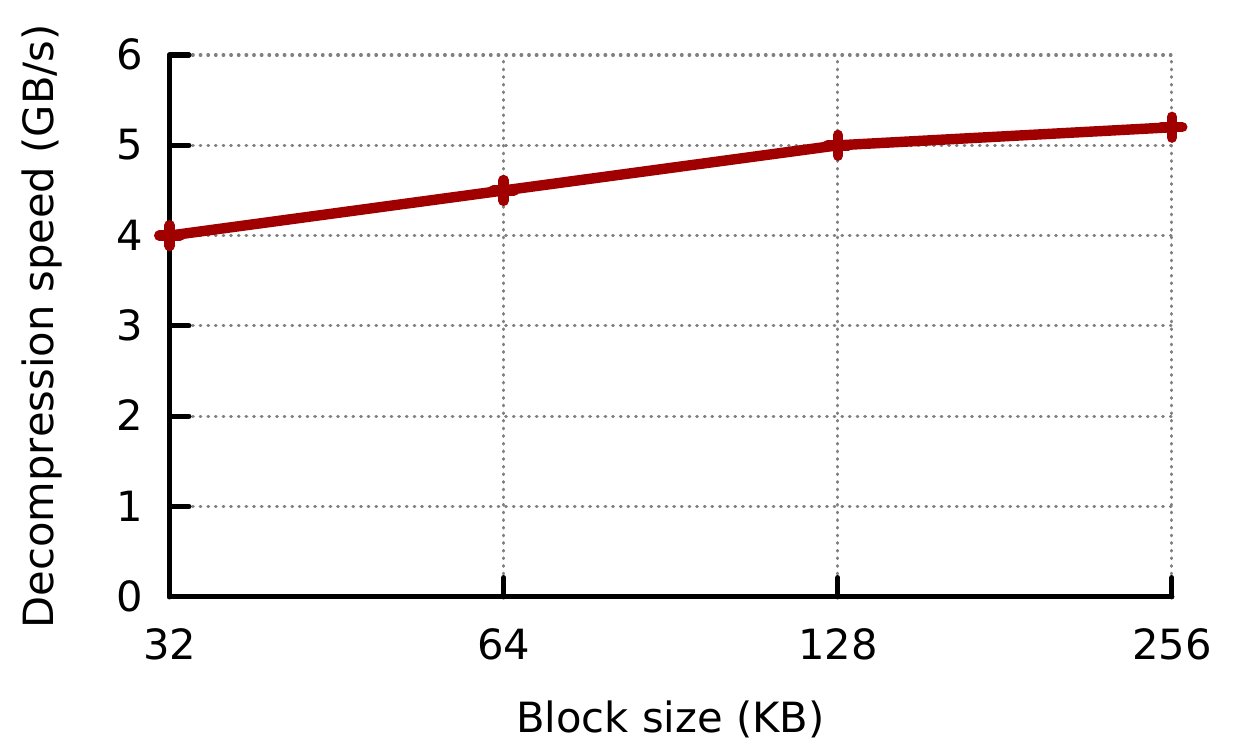}
  \includegraphics[width=0.8\linewidth]{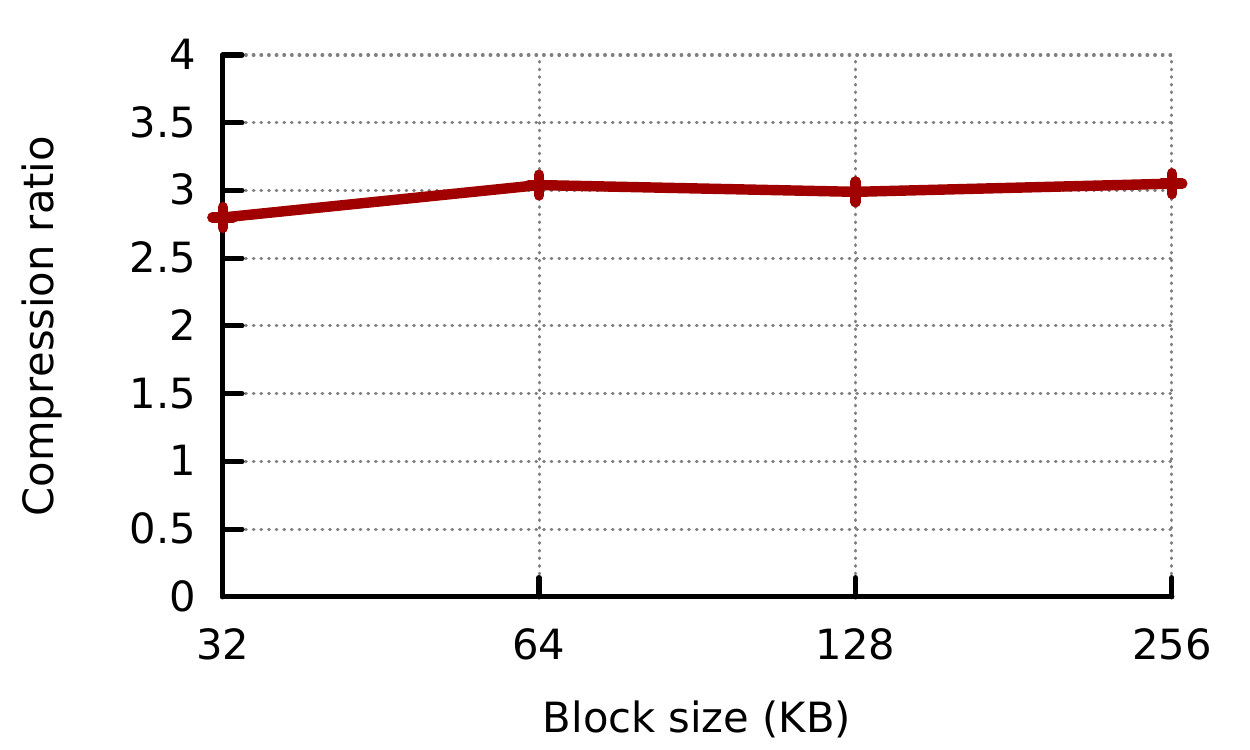}
  \caption{Decompression speed (data transfer cost included) and ratio of 
	\textbf{Gompresso/Bit} for different block sizes}
  \label{fig:decspeed_bsize}
\end{figure} 

\subsection{Dependency on Data Block Size} 
Figure~\ref{fig:decspeed_bsize} shows the decompression speed and compression
ratio for different data block sizes. Larger blocks increase the available
parallelism for Huffman decoding because there are more parallel sub-blocks in
flight. Threads operating on sub-blocks that belong to the same data block
share the Huffman decoding tables, which are stored in the software-controlled, 
on-chip memory of the GPU. This intra-block parallelism leads to a 
better utilization of the GPU's compute resources by scheduling more data 
blocks on the GPU's processors for concurrent execution (inter-block
parallelism). The space required by the Huffman decoding tables in the 
processors' on-chip memory limit the number of data blocks that can 
be decoded concurrently on a single GPU processor. 

Each Huffman decoding table has $2^{\mathrm{CWL}}$ entries, where $\mathrm{CWL}$
is the maximum codeword length.  To fit the look-up tables in the on-chip memory,
we are using limited-length Huffman encoding with a maximum length of
$\mathrm{CWL}=10\,\textrm{bits}$. Figure~\ref{fig:decspeed_bsize} shows that the 
compression ratio only marginally degrades for smaller blocks, so the space 
overhead of storing the block header for each compressed data block is not significant.

\begin{figure}[ht!] 
\centering
\begin{minipage}{.49\textwidth}
  \includegraphics[width=0.99\linewidth]{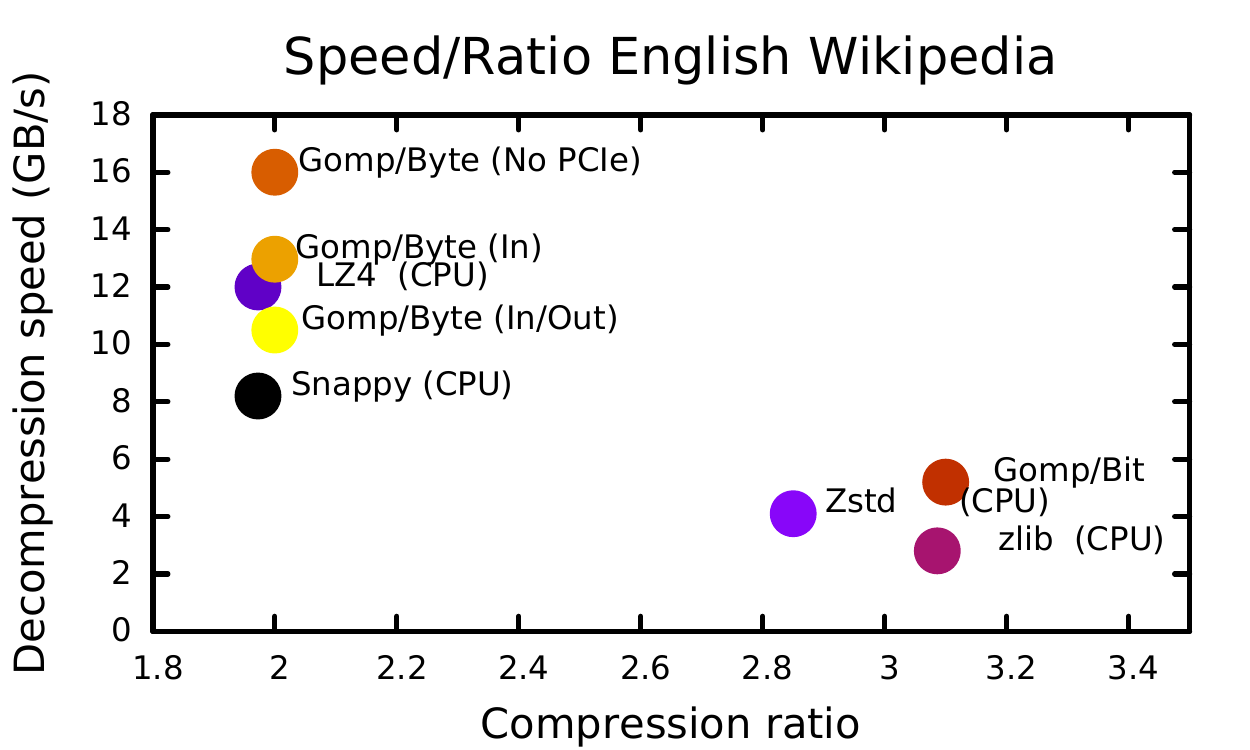}
\end{minipage}
\begin{minipage}{.49\textwidth}
    \includegraphics[width=0.99\linewidth]{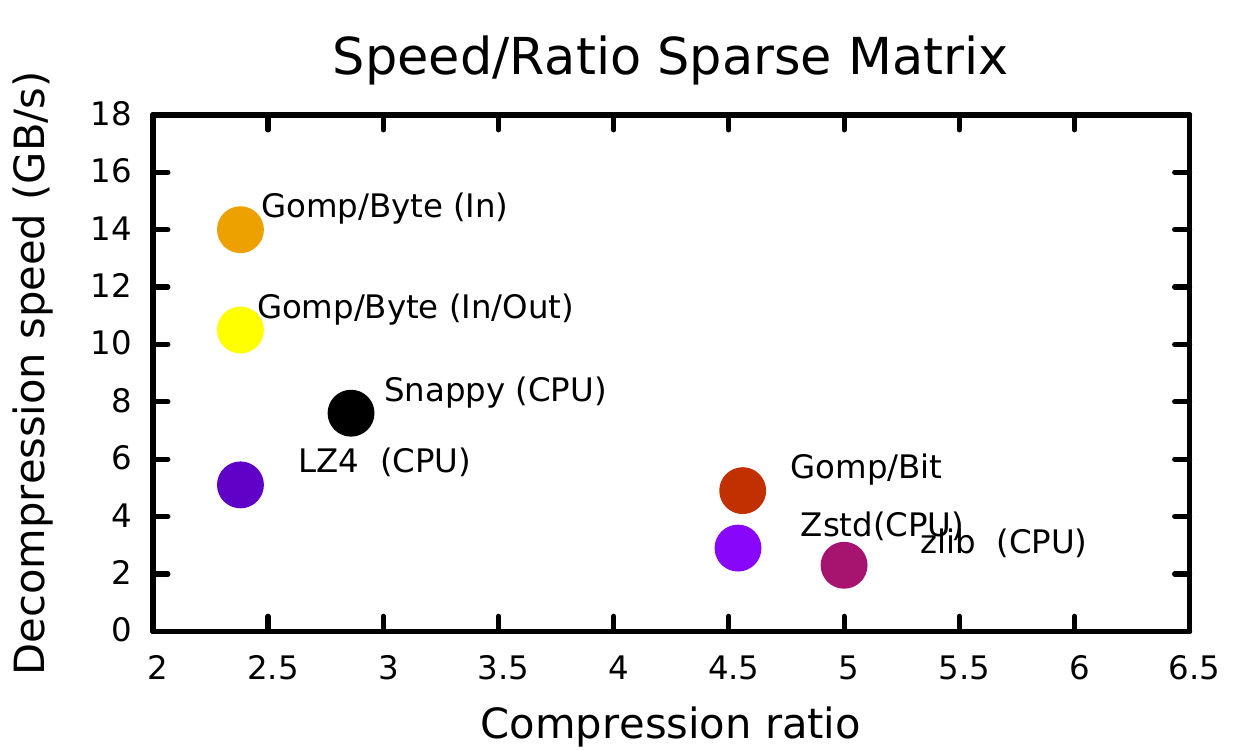}
\end{minipage}
 \caption{GPU vs multicore CPU performance. Cost for transferring data to and from the GPU 
is included for \textbf{Gompresso/Bit}. For \textbf{Gompresso/Byte},
 we show the performance both including and not including data transfers. }
   \label{fig:2dperf1}
\end{figure} 

\subsection{GPU vs. Multi-core CPU Performance} 
Lastly, we compare the performance of \textbf{Gompresso} to state-of-the-art parallel CPU
libraries regarding decompression speed and overall energy consumption. We used
a power meter to measure energy consumption at the wall socket. For CPU-only
environments, we physically removed the GPUs from our server to avoid including
the GPU's idle power. We parallelized the single-threaded implementations of the CPU-based state-of-the-art compression libraries by splitting the input data into equally-sized blocks that are then processed by the different cores in parallel. We chose a block size of 2\,MB, as this size resulted in the highest decompression speeds for the parallelized libraries. Once a thread  has completed decompressing a data block, it immediately processes the next block from a common queue. This balances the load across CPU threads despite input-dependent processing times for the different data blocks.

Figure~\ref{fig:2dperf1} shows the trade-offs between decompression speed and
compression ratio. In addition to the measurements of our \textbf{Gompresso} system, we
include the performance of two byte-level compression libraries (LZ4, Snappy)
and for two libraries using bit-level encoding (gzip, zlib) for comparison. Zstd
implements a different coding algorithm on top of LZ-compression that is
typically faster than Huffman decoding, and we include it in our measurements
for completeness \cite{zstd}. zlib implements the DEFLATE scheme for the CPU.
For the GPU measurements, we show the end-to-end performance, including times for:
(a) both compressed input and uncompressed output over PCIe, marked
\emph{(In/Out)} in Figure~\ref{fig:2dperf1}; (b) only the input transfers, marked
as \emph{(In)}; and (c) ignoring data transfers altogether, marked as \emph{No
PCIe}. 

For \textbf{Gompresso/Byte}, PCIe transfers turned out to be the bottleneck. In separate
bandwidth tests, we were able to achieve a PCIe peak bandwidth of 13\,GB/sec.
\textbf{Gompresso/Bit}, though not PCIe-bound, is still $2\times$  faster than zlib and
\textbf{Gompresso/Byte} is $1.35\times$ faster than LZ4. For the matrix dataset, the
decompression speed of \textbf{Gompresso/Bit} is around $2\times$ faster than zlib.
There is around 9\,\% degradation in compression ratio because we use limited-length Huffman coding. 
Although it lowers the compression efficiency, it enables
us to fit more Huffman decoding tables into the on-chip memory. 

\begin{figure} 
\centering
 \includegraphics[width=1.0\linewidth]{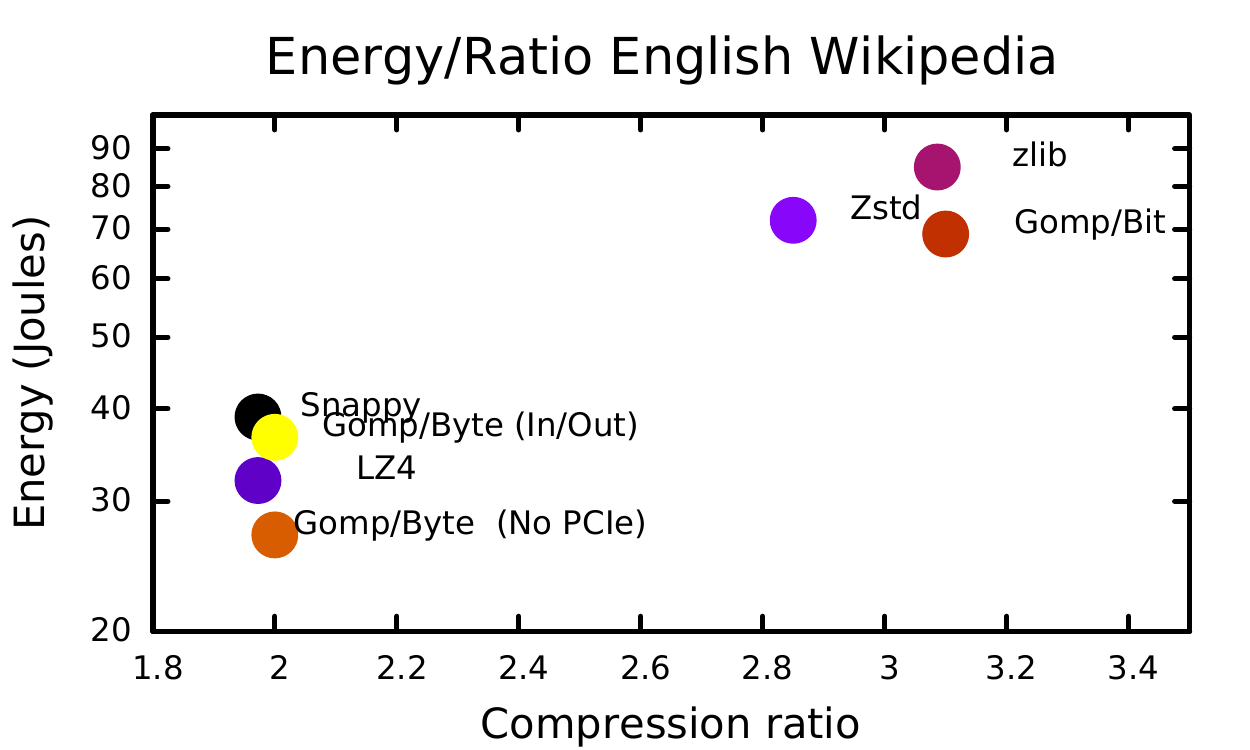}
\caption{GPU vs. multicore CPU energy consumption}
\label{fig:2denergy}
\end{figure} 

Finally, we compare the energy consumed to decompress the Wikipedia dataset.  In
general, faster decompression on the same hardware platform results in improved
energy efficiency. This is because the power drawn at the system level, i.e., at
the wall plug, does not differ significantly for different algorithms. More
interesting is the energy efficiency when comparing different implementations on
different hardware platforms, e.g., a parallel CPU vs. a GPU solution.
Figure~\ref{fig:2denergy} shows the overall energy consumption versus
the compression ratio for \textbf{Gompresso} and a number of parallelized CPU-based
libraries. \textbf{Gompresso/Bit} consumes 17\,\%  less energy than the
parallel zlib library. It also has similar energy consumption to Zstd, which
implements a faster coding algorithm.
